\documentclass[]{spie}  

\usepackage{amsmath,amsfonts,amssymb}
\usepackage{graphicx}
\usepackage[colorlinks=true, allcolors=blue]{hyperref}

\title{CCAT-prime: Designs and Status of the First Light 280 GHz MKID Array and Mod-Cam Receiver}

\author[a]{Cody J. Duell}
\author[a]{Eve M. Vavagiakis}
\author[b]{Jason Austermann}
\author[c]{Scott C. Chapman}
\author[a,d]{Steve K. Choi}
\author[e]{Nicholas F. Cothard}
\author[b]{Brad Dober}
\author[a]{Patricio Gallardo}
\author[b]{Jiansong Gao}
\author[f]{Christopher Groppi}
\author[d]{Terry L. Herter}
\author[a]{Zachary B. Huber}
\author[b]{Johannes Hubmayr}
\author[g,h]{Doug Johnstone}
\author[a,i]{Yaqiong Li}
\author[f]{Philip Mauskopf}
\author[j]{Jeff McMahon}
\author[a,d,i]{Michael D. Niemack}
\author[k]{Thomas Nikola}
\author[k]{Kayla Rossi}
\author[l]{Sara Simon}
\author[f]{Adrian K. Sinclair}
\author[d]{Gordon J. Stacey}
\author[b]{Michael Vissers}
\author[b]{Jordan Wheeler}
\author[e]{Bugao Zou}
\affil[a]{Department of Physics, Cornell University, Ithaca, NY, 14853, USA}
\affil[b]{Quantum Sensors Group, NIST, Boulder, CO, 80305, USA}
\affil[c]{Department of Physics and Atmospheric Science, Dalhousie University, Halifax, NS, B3H 4R2, Canada}
\affil[d]{Department of Astronomy, Cornell University, Ithaca, NY, 14853, USA}
\affil[e]{Department of Applied Physics, Cornell University, Ithaca, 14853, NY, USA}
\affil[f]{School of Earth and Space Exploration, Arizona State University, Tempe, AZ, 85281, USA}
\affil[g]{National Research Council, Herzberg Astronomy and Astrophysics, Victoria, BC, V9E 2E7, Canada}
\affil[h]{Department of Physics and Astronomy, University of Victoria, Victoria, BC, V8P 5C2, Canada }
\affil[i]{Kavli Institute at Cornell for Nanoscale Science, Cornell University, Ithaca, NY, 14853, USA}
\affil[j]{Department of Astronomy and Astrophysics, University of Chicago, Chicago, IL, 60637, USA}
\affil[k]{Cornell Center for Astrophysics and Planetary Sciences, Cornell University, Ithaca, NY, 14853, USA}
\affil[l]{Fermi National Accelerator Laboratory, Batavia, IL, 60510, USA}
\authorinfo{Further author information: (Send correspondence to C.J.D.) E-mail: cjd259@cornell.edu, Telephone: 1 607 255 0833}

\begin{document} 
\maketitle

\begin{abstract}
The CCAT-prime project’s first light array will be deployed in Mod-Cam, a single-module testbed and first light cryostat, on the Fred Young Submillimeter Telescope (FYST) in Chile’s high Atacama desert in late 2022. FYST is a six-meter aperture telescope being built on Cerro Chajnantor at an elevation of 5600 meters to observe at millimeter and submillimeter wavelengths\cite{Stacey}. Mod-Cam will pave the way for Prime-Cam, the primary first generation instrument, which will house up to seven instrument modules to simultaneously observe the sky and study a diverse set of science goals from monitoring protostars to probing distant galaxy clusters and characterizing the cosmic microwave background (CMB). At least one feedhorn-coupled array of microwave kinetic inductance detectors (MKIDs) centered on 280 GHz will be included in Mod-Cam at first light, with additional instrument modules to be deployed along with Prime-Cam in stages. The first 280 GHz detector array was fabricated by the Quantum Sensors Group at NIST in Boulder, CO and includes 3,456 polarization-sensitive MKIDs. Current mechanical designs allow for up to three hexagonal arrays to be placed in each single instrument module. We present details on this first light detector array, including mechanical designs and cold readout plans, as well as introducing Mod-Cam as both a testbed and predecessor to Prime-Cam.  
\end{abstract}


\section{INTRODUCTION}
\label{sec:intro}

The Fred Young Submillimeter Telescope (FYST) is a six-meter, off-axis, crossed-Dragone telescope offering a wide field-of-view and high-throughput that is currently being built by the CCAT-prime collaboration\footnote{CCAT-prime is an international consortium including researchers from the USA, Canada, Germany, and Chile.} to observe at millimeter and submillimeter wavelengths\cite{Stacey, aravena2019ccatprime}. FYST (pronounced ``feast") is being built near the summit of Cerro Chajnantor at an elevation of 5600 meters in the Atacama Desert of northern Chile. First light is expected  in late 2022. Planned broadband, polarimetric surveys at five different frequency bands (220, 280, 350, 410, and 850 GHz) along with simultaneous spectroscopic surveys (with $R\sim100$ from 210 to 420 GHz) will take advantage of FYST's combination of wide field-of-view, a low emissivity telescope, and extraordinary atmospheric conditions\cite{Stacey, aravena2019ccatprime}.

 \begin{figure} [ht]
   \begin{center}
   \begin{tabular}{c} 
  \includegraphics[width=0.4\textwidth]{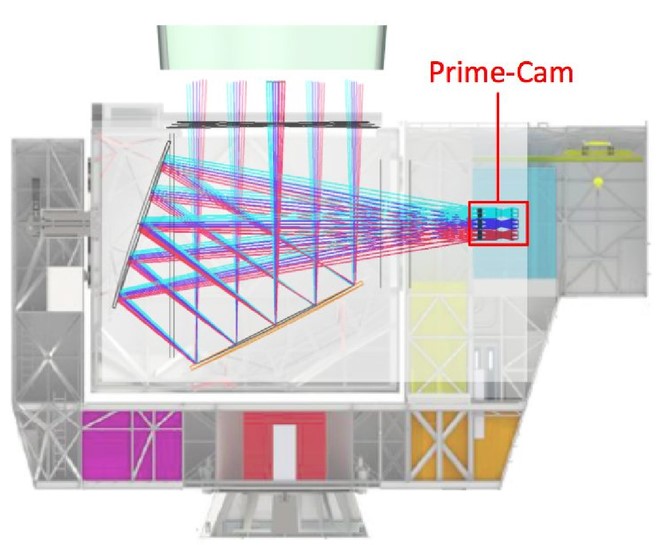}
  \includegraphics[width=0.4\textwidth]{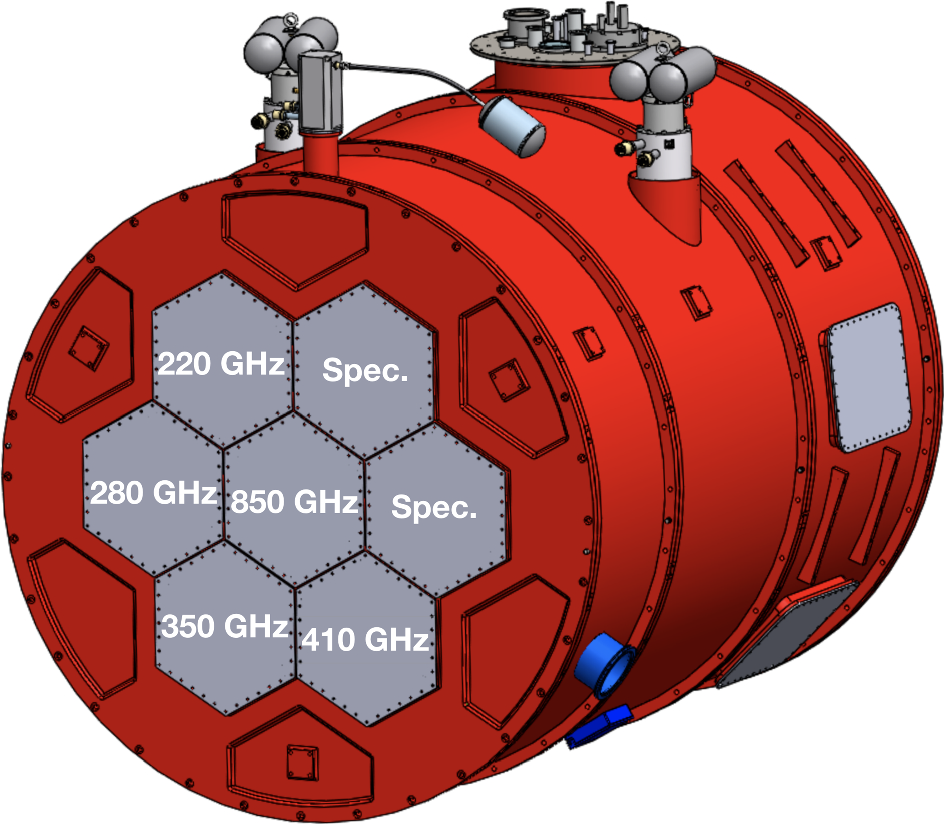}
  \label{fig:CCAT_PrimeCam} 
   \end{tabular}
   \end{center}
\caption{(Left) A cross-section of FYST with optics focused into the receiver cabin\cite{EMV2018}. (Right) A model of the Prime-Cam cryostat with a potential instrument module configuration\cite{choi_sensitivity_2020}.}
\end{figure}

With FYST's unparalleled survey capabilities in the submillimeter, CCAT-prime will target a diverse set of science goals in cosmology and far-infrared astronomy\cite{aravena2019ccatprime}. These include:

\begin{itemize}
  \item Investigating the formation, growth, and large-scale structure of the first star-forming galaxies through spectroscopic intensity mapping of the red-shifted [CII] line;
  \item Improving constraints on primordial gravitational waves and new particle species obtainable from observations of the cosmic microwave background (CMB) by characterizing signal-limiting foreground dust polarization across multiple wavelengths;
  \item Probing fundamental physics such as dark energy and the sum of the neutrino masses through the Sunyaev-Zel’dovich (SZ) effect;
  \item Revealing the effects of active galactic nuclei-star formation feedback in clusters by measuring the SZ signal for more than 1000 galaxy clusters;
  \item Tracing the history of dusty star formation by combining photometric measurements from CCAT-prime surveys with those made at optical and near-infrared wavelengths.
\end{itemize}

Much of the first-generation science goals will be tackled by Prime-Cam (shown in Figure 1), an instrument that has been previously detailed in Vavagiakis et. al. (2018) [\citenum{EMV2018}], which is capable of holding up to seven independent instrument modules for simultaneous observations\footnote{Additionally, a two-color heterodyne array receiver, CHAI (the CCAT-prime Heterodyne Array Instrument) will occupy 25\% of FYST's observing time over the first five years of operation.}. However, as Prime-Cam is unlikely to be ready at first light, the first 280 GHz array will be tested and deployed within Mod-Cam, a single-module testbed and first light instrument. The initial array will capitalize on advances in the fabrication of large format arrays of background-limited polarimeters\cite{Hubmayr,austermann_large_2018,austermann_millimeter-wave_2018} to deliver nearly 3,500 feedhorn-coupled, polarization-sensitive detectors on a single 15\,cm wafer and will lay the foundation for the eventual 100,000+ detectors that will be deployed on Prime-Cam. This first light array is described here in detail along with a description of Mod-Cam.

\section{First Light Detector Array}

\subsection{Detectors}

While the full deployment of Prime-Cam will be able to hold up to seven independent instrument modules with three detector arrays in each, at first light and for commissioning the telescope we will deploy Mod-Cam's singular instrument module with a single 280\,GHz array. Additional detector arrays and instrument modules are under development for deployment alongside the 280\,GHz array or shortly following in Prime-Cam, including an additional broadband module centered on 850 GHz and the spectrometer module EoR-Spec\cite{cothard_design_2020}. 

As with all arrays currently planned for CCAT-prime, the first light array will use microwave kinetic inductance detectors (MKIDs). An MKID is a superconducting resonator that derives a significant fraction of its total inductance from the kinetic inductance of an absorbing strip\cite{mazin, day_broadband_2003}. As photons strike the absorbing element of the detector, they break Cooper pairs and create quasiparticles, causing a change in the inductance and a resultant shift in the resonant frequency and quality factor. 

 \begin{figure} [ht]
   \begin{center}
   \begin{tabular}{c} 
  \includegraphics[height=4cm]{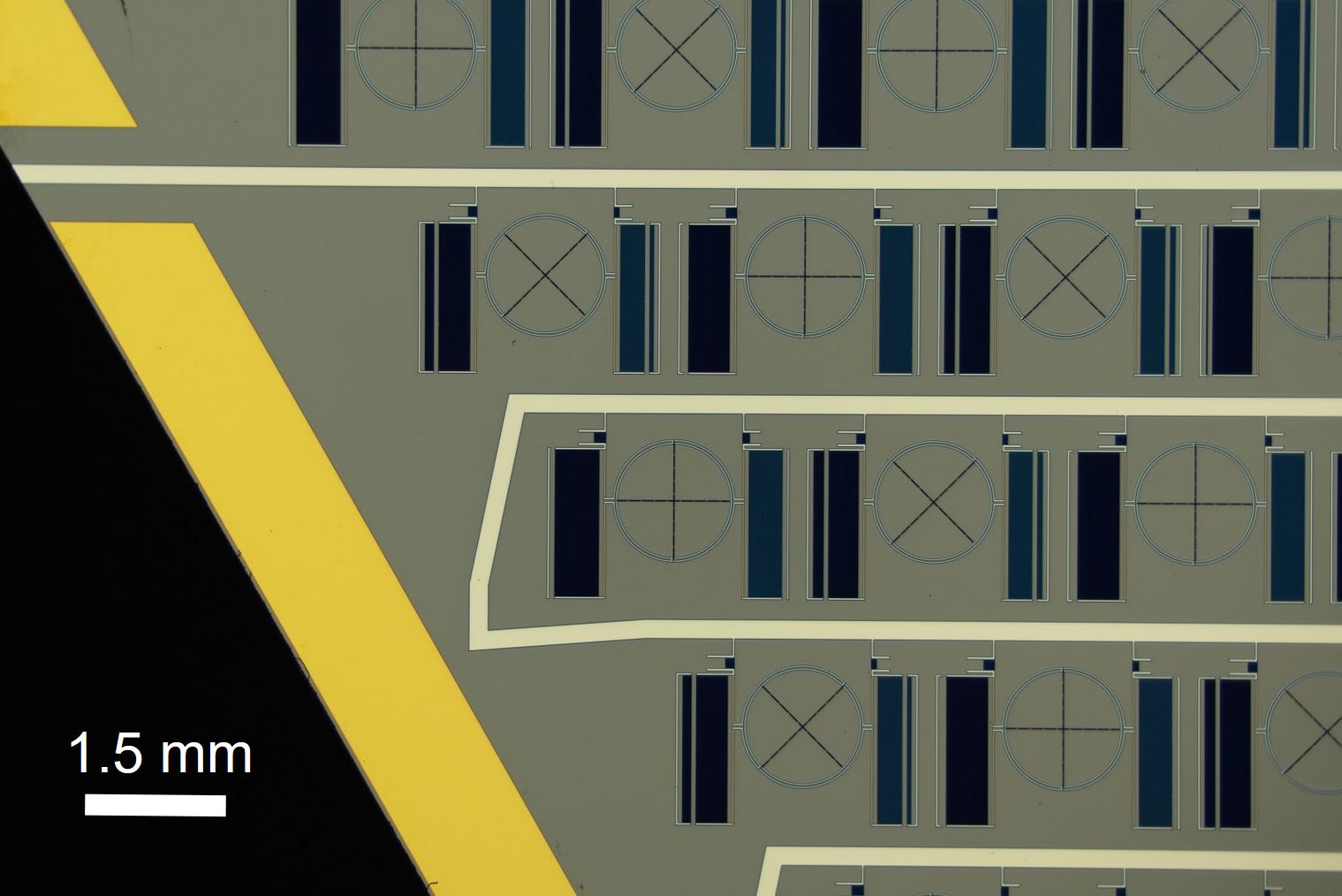}
  \includegraphics[height=4cm]{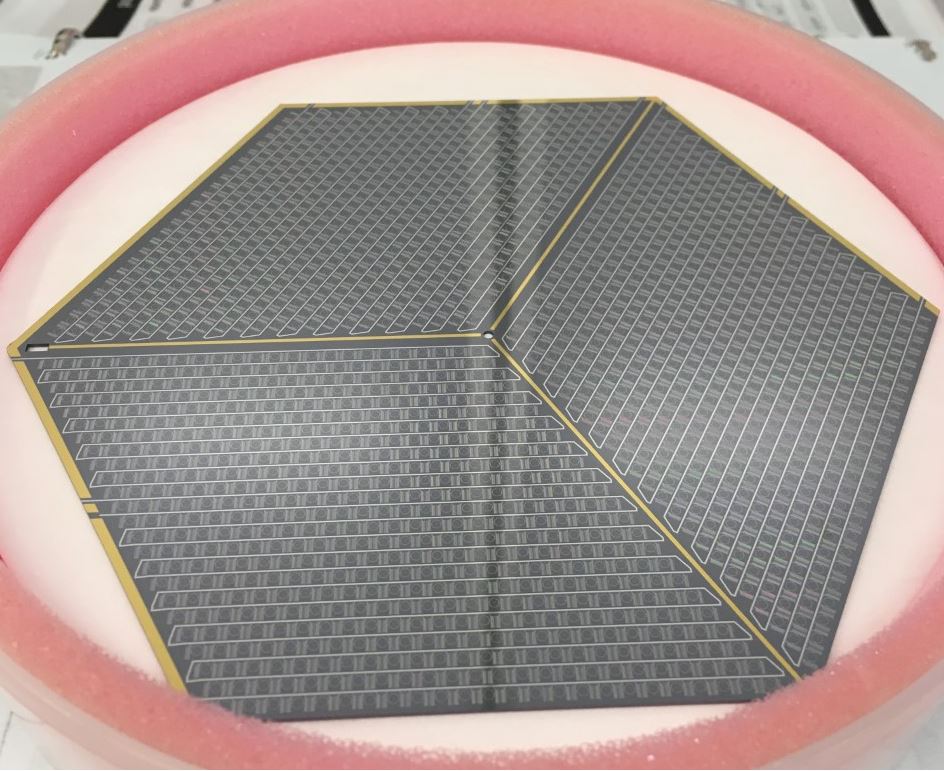}
  \includegraphics[height=4cm]{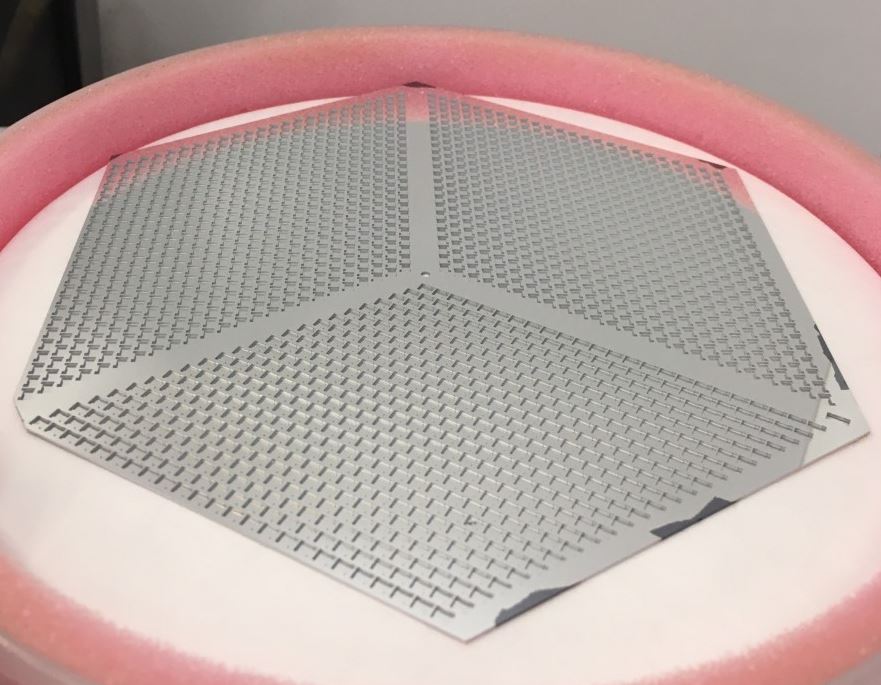}
   \label{fig:Detectors} 
   \end{tabular}
   \end{center}
\caption{(Left) Close up view of several pixels from the first light MKID array. (Center) Top of the completed first light 280 GHz array, which is approximately 13\,cm wide. (Right) Bottom of the first light array.}
\end{figure}

The initial array (seen in Figure 2) contains 3,456 feedhorn-coupled, polarization-sensitive MKIDs fabricated from TiN on a hexagonal 550-${\mu}$m thick, 15\,cm diameter silicon-on-insulator wafer. It is optimized for observing a $\sim$60-GHz wide band centered at 280 GHz with background-limited sensitivity\cite{choi_sensitivity_2020}. Fabrication of the first array was recently completed by the Quantum Sensors Group at the National Institute of Standards and Technology (NIST) in Boulder, CO. They were able to draw heavily on the experience gained through work on detectors for the BLAST-TNG\cite{dober_optical_2016, Galitzki} and TolTEC\cite{austermann_millimeter-wave_2018, austermann_large_2018} receivers. The resonators share the same design as the 280 GHz detectors designed for TolTEC, with minor adjustments in the absorber geometry to account for CCAT-prime's slightly lower atmospheric loading.

\subsection{Focal Plane Assembly Mechanical Designs}

The detector array is being mounted within a focal plane assembly that also holds the aluminum-machined feedhorns. This mechanical assembly (shown in Figure 3) serves to set the alignment between the detectors and the feedhorns, couple the detectors with the RF lines for readout, and provide heatsinking to the dilution refrigerator so as to keep the entire assembly stable at the detectors' 100 mK operating temperature. The hexagonal design allows for packing three arrays within a single instrument module, keeping all three as near as possible to the center of the instrument's focal plane. Designing and machining the mechanical components to meet the relatively strict alignment tolerances while minimizing risk of damaging the detector wafer during cooldowns provided several significant challenges.

 \begin{figure} [ht]
   \begin{center}
   \begin{tabular}{c} 
  \includegraphics[width=0.5\textwidth]{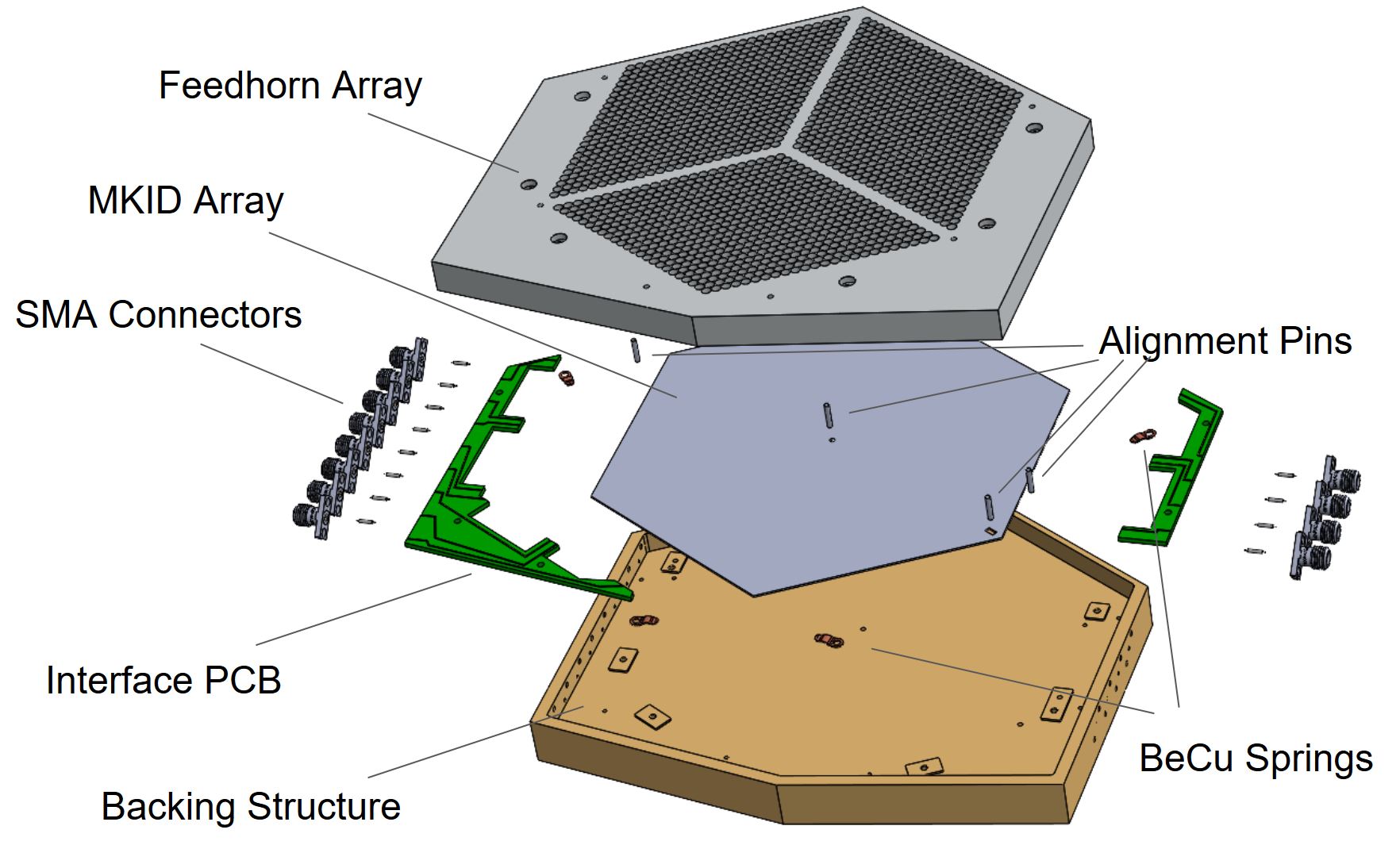}
  \includegraphics[width=5.5cm,height=4.2cm]{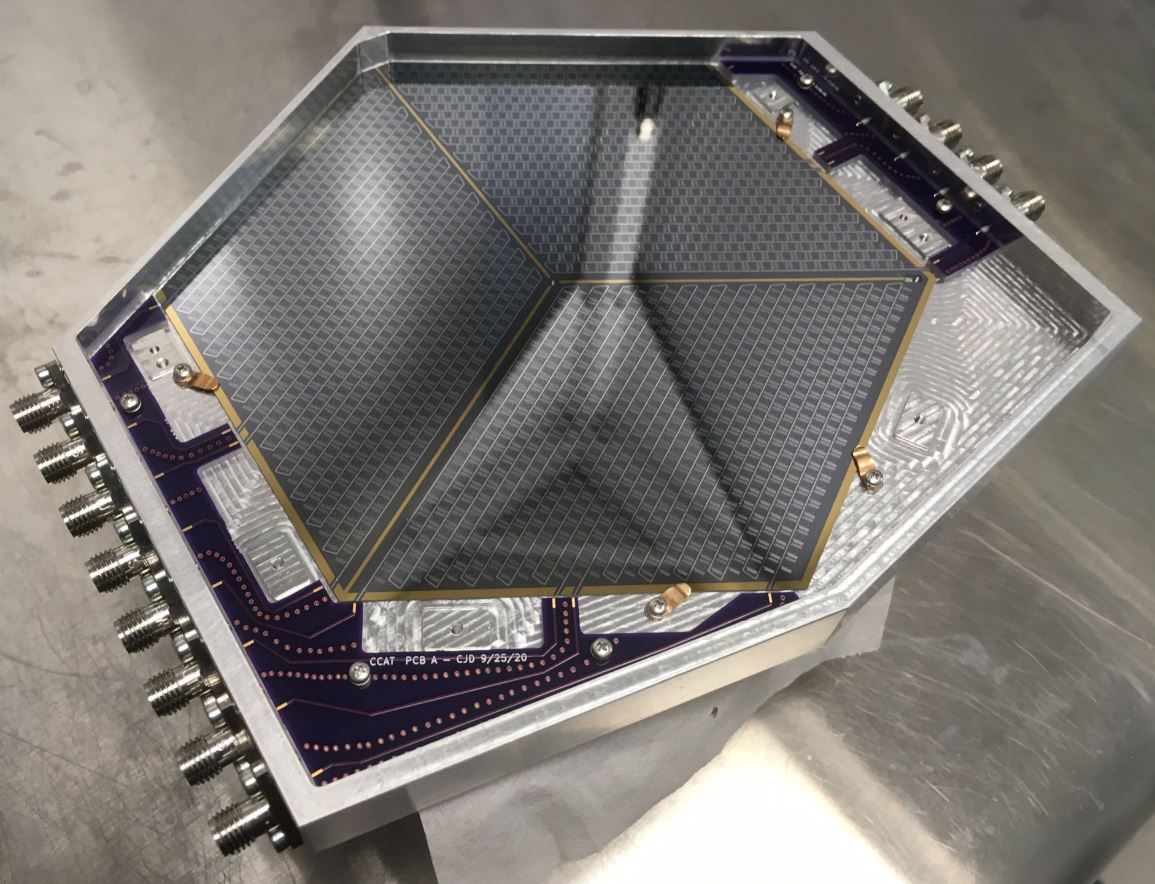}
  \label{fig:MechExp} 
   \end{tabular}
   \end{center}
\caption{(Left) An exploded view of the focal plane assembly, including all alignment pins and readout hardware, but with screws and pogo pins removed. (Right) The current status of the test assembly with a mechanical test wafer.}
\end{figure}

Detector alignment is achieved by use of a pin-and-slot design, combining a tightly-fitted central pin with a radial slot and pin to allow for differential thermal contraction between the silicon wafer and the gold-plated aluminum backing structure. The aluminum-machined feedhorn array is aligned to the detector wafer by means of a separate pair of tight-fitting pins along the outer edge of the detector array. Particular care was paid to minimizing potential strain on the silicon wafer generated by thermal gradients between the aluminum backing structure and the feedhorn array during the cooling process. Hence, the choice was made to align the detectors and feedhorns to the backing structure through separate alignment pins. When cold, a ~75 $\mu$m gap is achieved between the feedhorns’ choke structures and the detector wafer across the 150\,mm wafer by means of a set of raised mounting platforms along the outer edge of the detector wafer, which also serve as mounting points for the feedhorn array. These platforms, as well as the positions of the alignment pins, are shown in Figure 4.

Pogo pins are placed along designated lanes that spread out radially from the center of the detector array to reduce microphonics, sliding along a gold layer as the assembly cools. These lanes can be seen clearly in Figure 2. An additional layer of gold is placed along the edges of the detector array on four of the six sides allowing for gold wirebonds between the wafer's edge and the backing structure. This provides additional heat sinking for the array, while also improving the grounding beyond just through surface contact with the backing structure. 

The feedhorn array, including all choke structures, is being machined out of aluminum based on a spline profile\cite{austermann_millimeter-wave_2018}. To reduce the potential for warping of the feedhorn array during thermal cycling due to built-up stress from machining, several rounds of thermal annealing were employed following an initial rough machining of the array. The annealing process involved rapidly moving the array between baths of liquid nitrogen ($\sim$77 K) and boiling water ($\sim$373 K) several times. This should significantly reduce stress in the feedhorns, and thereby reduce the possibility of damaging the array during cooldowns or warm-up.

  \begin{figure} [ht]
  \begin{center}
  \begin{tabular}{c} 
  \includegraphics[width=0.7\textwidth]{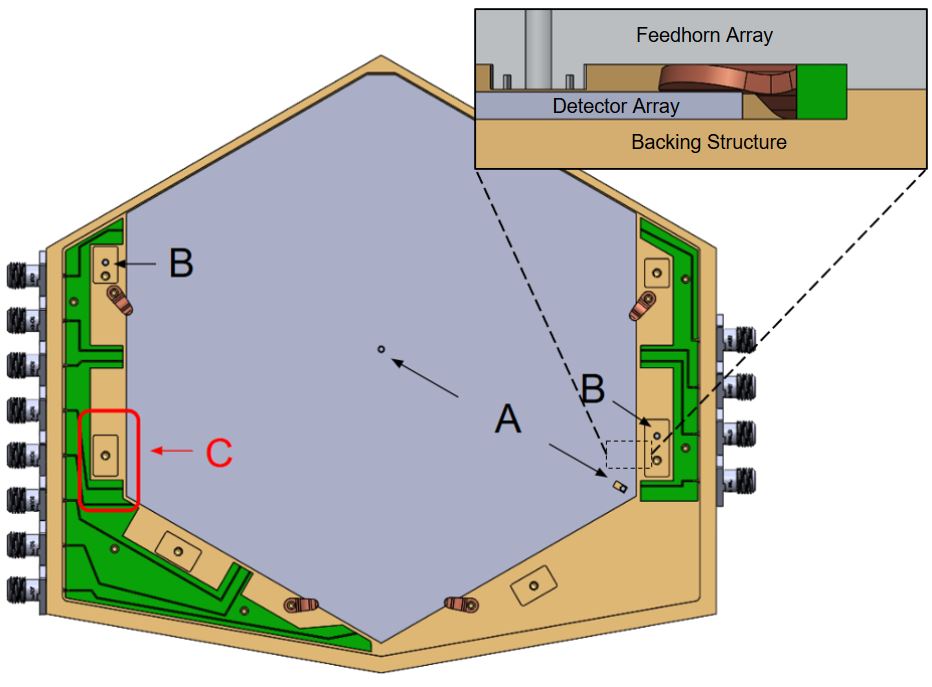}
  \label{fig:FrontOfMech} 
   \end{tabular}
   \end{center}
   \caption{Several key features of the mechanical designs are labeled. (A) The pin-and-slot feature for detector alignment. (B) Separate pins for aligning the aluminum-machined feedhorn array on the aluminum base. (C) One of six raised platforms that help to set a ~75 $\mu$m cold gap between the detectors and feedhorn array. and serve as mounting points for the feedhorn array. A close-up cross-sectional view of one of these is shown in the top right.}
\end{figure}

\subsection{Readout}

The first light array is split into 6 networks, each requiring 576 resonators with resonant frequencies designed to run between 500 MHz and 1 GHz to be measured over a single RF feedline. One of the major advantages of using MKIDs for the first light array is the relative simplicity of their readout. Since they are fundamentally resonators, MKIDs are naturally frequency-multiplexed by tuning the individual resonator parameters during fabrication. Thus the majority of the readout complexity is confined to the room temperature electronics, while the cryogenic components are primarily comprised of RF lines, attenuators, and low noise amplifiers built at Arizona State University. 

Room temperature readout electronics for both lab-based testing and full-scale deployment will feature a reconfigurable FPGA-based design. Our readout approach is evolving from the second generation Reconfigurable Open Architecture Computing Hardware (ROACH-2), including firmware developed for BLAST-TNG\cite{GordonBLAST} and TolTEC. This system uses a ROACH-2 FPGA board coupled with a DAC/ADC board for generating and capturing probe tones, as well as data streaming. A baseband frequency comb is generated in these digital elements and sent through an analog front-end for IQ mixing and up-conversion to the required band. The frequency comb is then sent through the cryogenic readout components to probe the detectors, passing again through the analog front-end for down-conversion before being sent to the ADCs. 

While this readout system is seeing successful use by BLAST-TNG and TolTEC, it presents significant challenges in the long-term as we look towards the full deployment of Prime-Cam with up to 21 KID arrays. With this single array requiring six separate ROACH-2 systems for its six total networks, it would prove challenging to readout the entirety of Prime-Cam with ROACH-2 systems due to space and power considerations\footnote{Each individual ROACH-2 system requires roughly two units of rack space and 100 W of power.}, even if we were to see significant improvements in the multiplexing factors. With that in mind, work is ongoing towards porting the ROACH-2 firmware to the Xilinx ZCU111 RFSoC evaluation board\cite{sinclair}. This would enable significant reductions in the hardware requirements for readout by sharing digital signal processing resources between as many as eight feedlines.

\section{Mod-Cam: Testbed and First Light Instrument}

   \begin{figure} [ht]
   \begin{center}
   \begin{tabular}{c} 
  \includegraphics[width=0.53\textwidth]{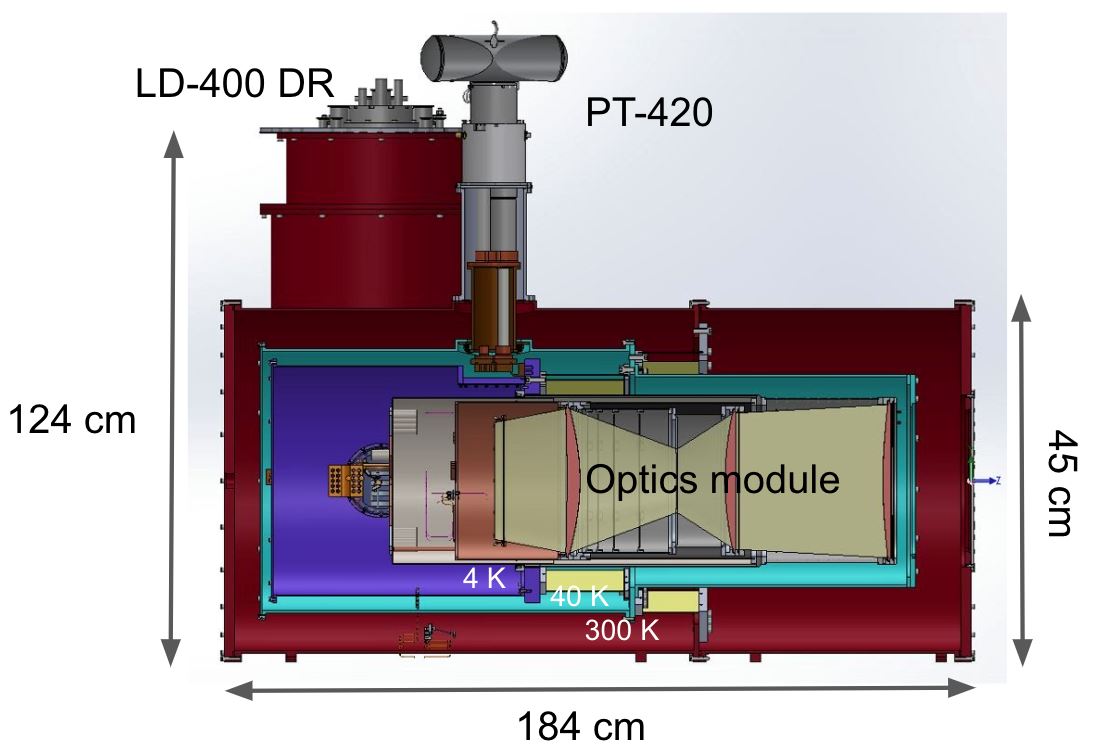}
  \includegraphics[width=0.43\textwidth]{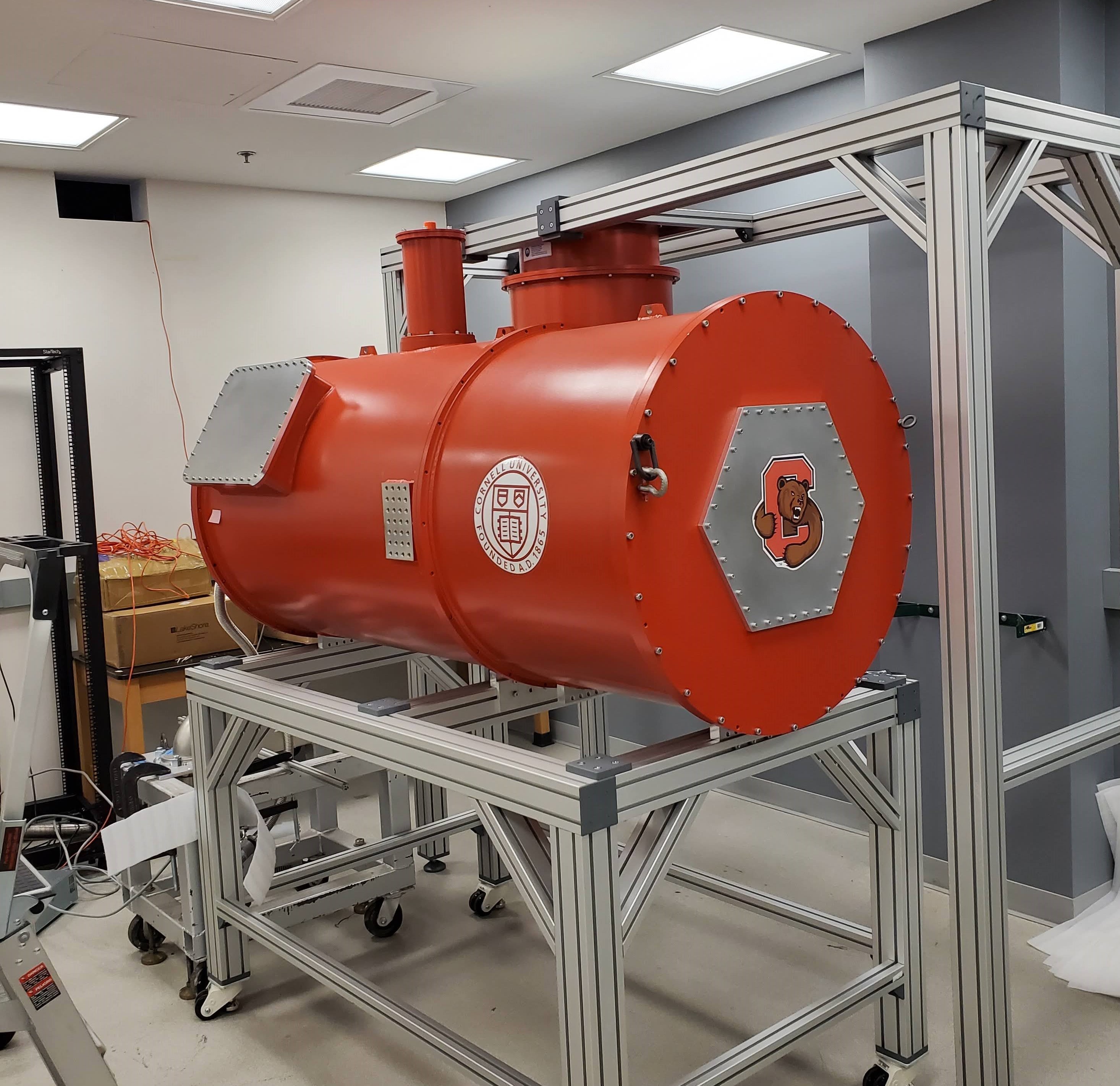}
   \end{tabular}
   \end{center}
   \label{fig:ModCamFig}
   \caption{(Left) Labeled rendering of Mod-Cam with side-car dilution refrigerator configuration shown. (Right) Current status of Mod-Cam at Cornell.}
\end{figure}

The first array will be tested and deployed in Mod-Cam, a single-module cryogenic testbed for Prime-Cam\cite{EMV2018,choi_sensitivity_2020} and first light instrument for CCAT-prime (seen in Figure 5). Mod-Cam's 45-cm diameter cryostat is cooled by a Bluefors LD-400 dilution refrigerator (DR) and enables efficient swapping of instrument modules by means of its side-car DR design. The instrument modules are installed from the back of the cryostat and are cantilevered off of the 4 K stage. Each instrument module tested or deployed in Mod-Cam will be optimized for a specific subset of the overall science goals and be able to hold up to three 100 mK detector arrays along with silicon lenses and filter stacks at 1 K and 4 K.

The modules themselves allow for up to a 36-cm diameter aperture and are based on the optics tube designs for the Simons Observatory's large aperture telescope receiver\cite{EMV2018, Zhu, dicker}. In this design, light enters the optics module after passing through the 300 K ultra-high-molecular-weight polyethylene (UHMWPE) vacuum window and 40 K infrared-blocking filters. The light is additionally filtered by a series of absorbing alumina filters\cite{dicker}, metal-mesh infrared-blocking filters \cite{tucker}, and low pass edge filters to block unwanted radiation, and it is re-imaged onto the focal plane by three meta-material anti-reflection-coated silicon lenses\cite{Datta:13}.

Mod-Cam is designed to serve as a scaled-down version of the much larger Prime-Cam for significantly faster testing of individual instrument modules prior to deployment. It has a 45-cm diameter exterior aluminum vacuum shell, along with additional aluminum shells at 40 K and 4 K supported by a series of G10 tabs. In addition to the side-mounted DR, an optional Cryomech PT-420 pulse tube can provide cooling power at 40 K and 4 K. All thermometry and RF signals are read out through a custom modular harness that is installed on the opposing side to the DR. The modularity of the harness design allows for flexible and upgradable readout options. This arrangement is what leaves the rear of Mod-Cam relatively clear for removal of both individual detector arrays and entire instrument modules.

\section{Current Status}

The first-light detector array, which recently completed fabrication at NIST, is currently undergoing initial dark testing to measure resonant frequencies and quality factors. Preliminary measurements of a test pixel are shown in Figure 6, indicating the change in resonator performance in response to a cryogenic blackbody source. While this early data is promising, additional testing is required before any conclusions can be made about the initial detector yield or sensitivity. The full mechanical array, with the exception of the feedhorn array (which is currently in production), has been fabricated, assembled, and successfully undergone cryogenic testing with a mechanical test wafer. Additionally, all components of the Mod-Cam cryostat have been delivered, assembled, and leak-checked at Cornell. Currently, work is focused on fabricating the remaining mechanical components for the first instrument module and integrating the cryogenic components and thermometry for cryogenic tests.  

  \begin{figure} [ht]
  \begin{center}
  \begin{tabular}{c} 
  \includegraphics[width=0.45\textwidth]{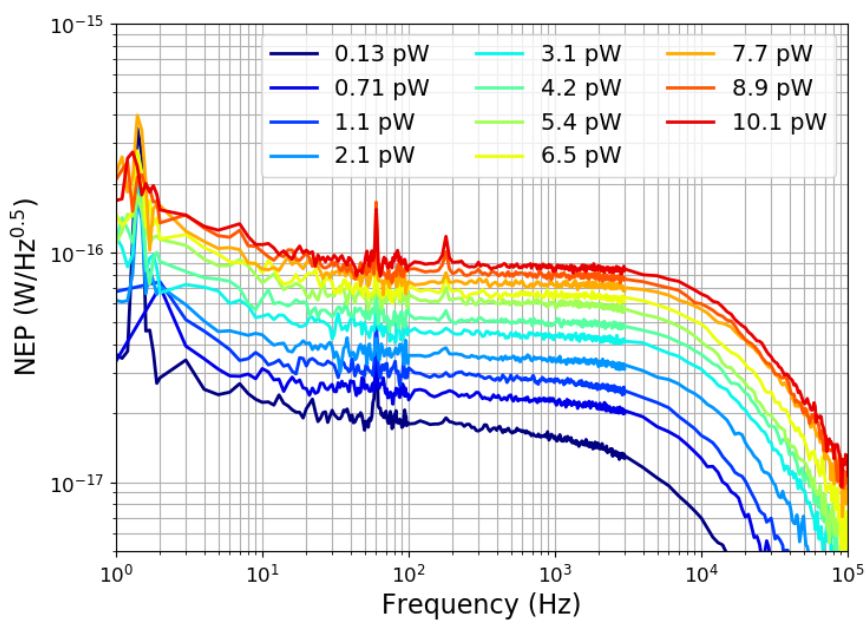}
  \includegraphics[width=0.45\textwidth]{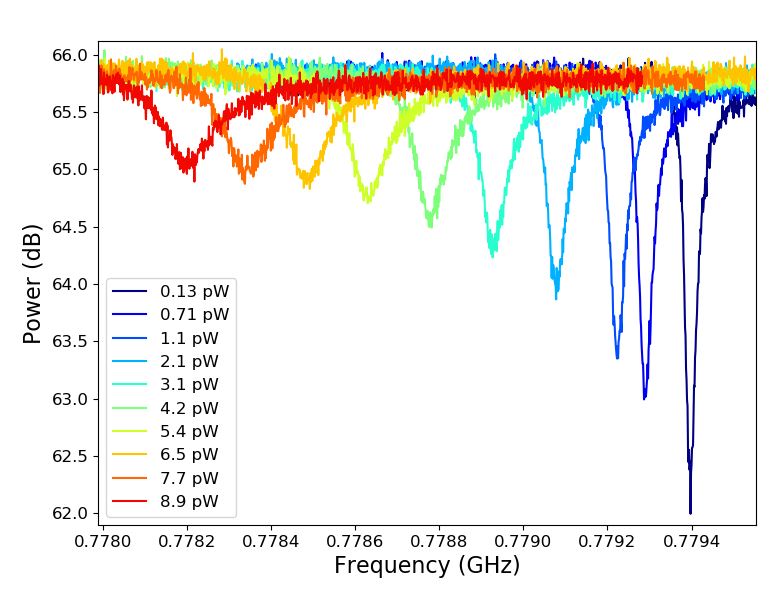}
  \label{fig:PrelimData} 
   \end{tabular}
   \end{center}
   \caption{Preliminary data taken at NIST showing the noise performance (left) and resonator response (right) of a recently completed test pixel as a function of loading power from a cryogenic blackbody. In both cases, optical power is roughly estimated due to uncertainties in the optical chain, and systematic noise sources have not been removed. }
\end{figure}

\section{Future Plans}

Upon receipt of the feedhorn array, we will finish validating the mechanical designs for the first array by cooling down the complete integrated test assembly, at which point the backing structure will be gold-plated and integration with the real detector wafer can take place. The unloaded resonant frequencies and quality factors of the first light array will be measured in the coming weeks, then full optical characterization will take place upon integration with the assembly at Cornell using a Fourier Transform Spectrometer and a cold load. Mod-Cam will soon undergo cryogenic tests in preparation for integration of the first light instrument module.

\section{Conclusion}

CCAT-prime's first light 280 GHz array and focal plane module is well on its way to completion for use on the upcoming Fred Young Submillimeter Telescope. With nearly 3,500 background-limited, polarization-sensitive MKIDs coupled to FYST's high throughput optics within Mod-Cam, it will pave the way for the more than 100,000 detectors that will eventually be deployed in Prime-Cam by the CCAT-prime collaboration. We presented here the detectors, mechanical designs, and readout plans for this first light array, along with an introduction to and overview of Mod-Cam which will serve as a precursor to Prime-Cam and cryogenic testbed for all of Prime-Cam's instrument modules.  

\acknowledgments      
 
CCAT-prime funding has been provided by Cornell University, the Fred M. Young Jr. Charitable Fund, the German Research Foundation (DFG) through grant number INST 216/733-1 FUGG, the Univ. of Cologne, the Univ. of Bonn, and the Canadian Atacama Telescope Consortium. EMV acknowledges support from the NSF GRFP under Grant No. DGE-1650441. MDN acknowledges support from NSF award AST-1454881. NFC acknowledges support from a NASA Space Technology Research Fellowship. SKC acknowledges support from NSF award AST-2001866.  YL acknowledges support from the Kavli Institute at Cornell for Nanoscale Science.

Work supported by the Fermi National Accelerator Laboratory, managed and operated by Fermi Research Alliance, LLC under Contract No. DE-AC02-07CH11359 with the U.S. Department of Energy. The U.S. Government retains and the publisher, by accepting the article for publication, acknowledges that the U.S. Government retains a non-exclusive, paid-up, irrevocable, world-wide license to publish or reproduce the published form of this manuscript, or allow others to do so, for U.S. Government purposes.

\bibliography{report} 
\bibliographystyle{spiebib} 

\end{document}